# DESIGN OF A SMART EMBEDDED UNINTERRUPTED POWER SUPPLY SYSTEM FOR PERSONAL COMPUTERS

Adamu Murtala Zungeru[1], Ambafi James Garba[2], Jonathan Gana Kolo[1], Mahmud Shehu Ahmed[2] and Ijarotimi Olumide[3]

[1]School of Electrical and Electronic Engineering, University of Nottingham, Jalan Broga, 43500 Semenyih, Selangor Darul Ehsan, Malaysia
adamuzungeru@ieee.org, jgkolo@gmail.com
[2]Department of Electrical and Electronic Engineering, Federal University of Technology, Minna, Nigeria
mahmudmsa@yahoo.com, ambafi@yahoo.com
[3]Electrical and Electronic Engineering Technology, Rufus Giwa Polytechnic, Owo, Ondo, Nigeria
ijarotimiolumide@yahoo.com

## ABSTRACT

*Digital equipment such as computers, telecommunication systems and instruments use microprocessors that operate at high frequencies allowing them to carry out millions or even billions of operations per second. A disturbance in the electrical supply lasting just a few milliseconds can affect thousands or millions of basic operations. The result may be malfunctioning and loss of data with dangerous or costly consequences (e.g. loss of production). That is why many loads, called sensitive or critical loads, require a supply that is protected. Many manufacturers of sensitive equipment specify very strict tolerances, much stricter than those in the distribution system for the supply of their equipment, one example being Computer Business Equipment Manufacturer's Association for computer equipment against distribution system disturbances. The design of this uninterrupted power supply (UPS) for personal computer (PC) is necessitated due to a need for enhanced portability in the design of personal computer desktop workstations. Apart from its original functionality as a backup source of power, this design incorporates the unit within the system unit casing, thereby reducing the number of system components available. Also, the embedding of this unit removes the untidiness of connecting wires and makes the whole computer act like a laptop. Not to be left out is the choice of a microcontroller as an important part of the circuitry. This has eliminated the weight and space-consuming components that make up an original design. The singular use of this microcontroller places the UPS under the class of an advanced technology device.*

## KEYWORDS

*Embedded System, Uninterrupted Power Supply, Personal Computer, Automation, Power Electronics.*

## 1. INTRODUCTION

An uninterruptible power supply, commonly called a UPS is a device that has the ability to convert and control direct current (DC) energy to alternating current (AC) energy. It uses a conventional battery of 12V rating as the input source and by the action of the inverter circuitry, it produces an alternating voltage which is sent to the load. This particular UPS is designed for a small scale load like a personal computer and hence only a basic power rate is generated by the UPS. Many believe that because an inverter is operating from a nominal 12V battery and it cannot deliver as much output as a normal mains power outlet, it's relatively safe. This is not usually true. Even a low power inverter rated at a mere 60watts has an output which is potentially fatal if you become its load. Such an inverter can have a typical output of 350mA at 230V. This is above ten (10) times the current level connected to cause fatal fibrillation and stop your heart.

Generally, uninterrupted power supply (UPS) can be grouped by source or method of functionality. (1) By Source: Here we have a voltage source (DC) for its operation or a current source (DC). The current source however is used for very high power consumption devices hence this design is a voltage source UPS. (2) By Functionality: Amongst others here is the single-tracked and dual-tract UPS. The single-tract UPS feeds the load continuously from the rectified DC supply directly. This type of UPS is disadvantaged because a fault in the rectification stage leads to a complete system failure. The dual tract acts like the single tract but it has a bypass that sources from the mains supply. Hence the battery is used only as backup and does not run all the time unlike the single track. This design is a dual track methodology. For an ideal UPS, basic functionality is needed (1) Being a backup utility, a UPS must ensure that there is no break in the power supply at any point in time unless major faults like fuse cuts are experienced. (2) An ideal UPS must provide the battery with an adequate charge so as to maintain the optimum conversion rate to AC when needed. (3) It must also ensure overcharge protection to prevent the battery from being damaged. (4) All forms of surges and undesired waveforms that may emanate from inverted source voltage are to be filtered and well suited to the output level. (5) Must be sensitive to maintain stability when the battery safe voltage is being exceeded. (6) It must also provide an overload protection for the entire unit.

Many embedded devices provide a rich GUI-based user experience; use file systems, multiprocessing, and multi-threading; and include networking. An operating system (OS) can provide these features to support the rapid development of application programs [1, 2]. In charging a battery of the personal computer (PC), a cheap, unattended, unregulated charger can destroy a battery by overcharging it. A temperature compensated charger is also highly recommended [3]. Thus most power supplies have a PWM controller based on the well-known TL494 [4] or equivalent chips (for instance KA7500). TL494 features two error amplifiers, but most power supplies only use one of these. A PWM controller featuring two error amplifiers is recommended in some design because one controls the output voltage and the other controls the output current.

However, after careful consideration of any existing design of the UPS and some embedded systems, this particular design incorporated the following methodology upgrades: (1) The battery charging unit is basically handled by the micro-controller which detects in split microseconds the point at which the safe battery (voltage at which operating the battery to generate alternating voltage is not safe) is being exceeded. This causes the system to shut down in order to prevent damage to the battery. (2) Also handled by the microcontroller is the overcharge protection. The controller disengages the battery at full charge voltage. (3) Application software interfaced via the USB (Universal Serial Bus) port of the computer motherboard maintains a constant check link between the operating system and the UPS. (4) To enhance compactness, 2-pole relays and switches are used to eliminate duplication of components. (5) Very simple and readily available components are sourced making the device commercially viable.

For clarity and neatness of presentation, the article is outlined into five (5) sections. The First Section gives a general introduction of a UPS and smart embedded systems. Review of system components used for this system design is presented in Section Two. In Section Three, we outline the design and implementation procedures. Section Four presents the experimental results and discussion of the results. In Section Five, we conclude the work with some recommendations. Finally, the references are presented at the end of the paper.

## 2. REVIEW OF SYSTEM COMPONENTS

This section discusses the basic theory of components used for this work. Though, we will be more focused on the heart of the system design (Microcontrollers) and its peripherals while we leave other basic electronic components. But interested readers can see [5-9] for theory of other components used.

## 2.1. Relays

Relays are electromechanical devices or solid state devices which operate in response to a signal which may be voltage, current, temperature etc. Electromagnetic relays operate due to magnetic fields. They are composed basically of two parts: (1) The operating coil and (2) The magnetic switch. When an input pulse is introduced into the coil, a magnetic field is produced in the core of the electromagnet. This action causes the switch to slide. Relays are either normally open or normally close. Relays are available for DC or AC excitation and coil voltages range from 5V to 230V. The primary use of relays is in remote switching, whereby the circuit is to be switched is electrically isolated from the switching circuit. This system utilizes both one-pole and two-pole normally open 12V DC relays. As shown in Fig. 1 below is a single pole and double pole relays, in which (a) has a single pole single throw, and single pole double throw. (b) Has a double pole single throw and double pole double throw.

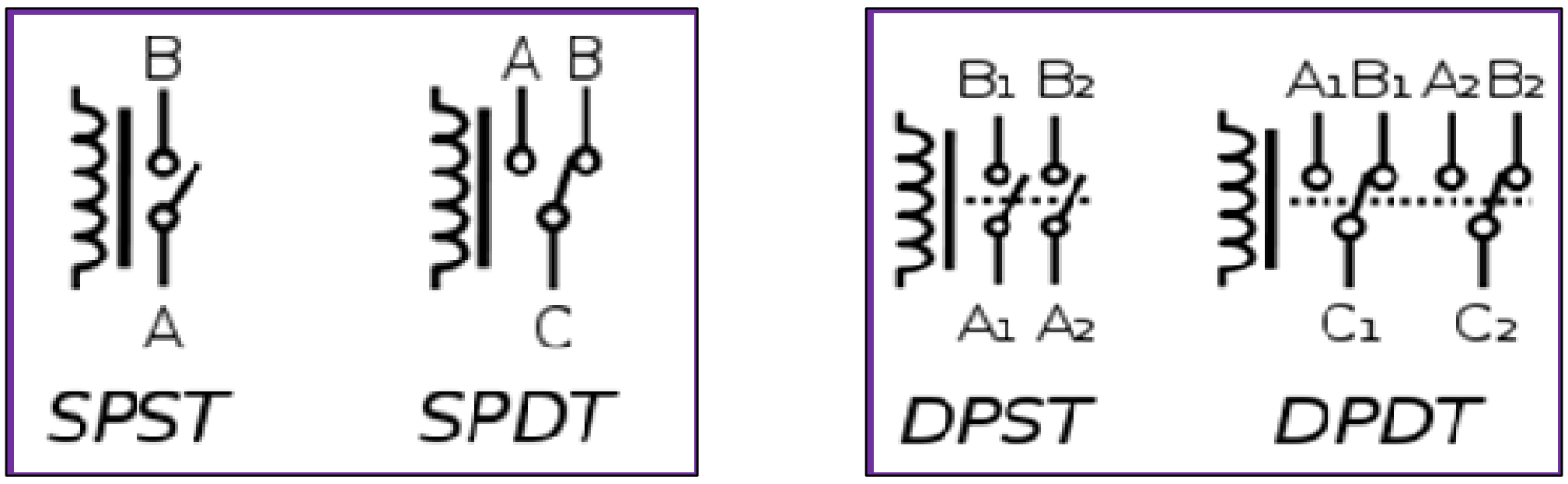

Fig. 1(a): One Pole-relay (b) Two Pole-relay

## 2.2. Microcontrollers

A microcontroller (MCU) is a single computer chip or integrated circuit that has the ability to execute written user programs. The MCU is normally used for the purpose of controlling some devices – this actually gives it its name microcontroller. The user program can be stored within the MCU or on an external chip called an Erasable Programmable Read Only Memory (EPROM).

MCU are normally integrated into small devices like the microwave ovens, keyboards and cell phones. The microprocessor that is universally accepted is not the same as a microcontroller. An MCU requires a small amount of computing power, less memory and very little attachment accessories. MCU-based systems are far more reliable and cheaper. Their small size also makes them desirable for circuit designers. The choice of MCU used in this system has twenty (20) pins and runs on a DC voltage of 5V. It possesses an internal comparator that acts like an OP-AMP comparator. It also has a clock (crystal) that runs at a frequency of 12MHz – this frequency is chosen so as to make the MCU trigger faster. The MCU takes charge of sending pulses that enable the charging circuit for the battery, the software application interface and the tracking of safe battery operational level.

### 2.2.1 Oscillator – Clock generator

A clock is used by the MCU to execute a program or sets of program instructions. To provide an MCU with a clock, an oscillator is used. There are different types of oscillators. Amongst others are the crystal oscillator and the resistor-capacitor pair.

#### 2.2.1.1 Crystal Oscillator

This consists of one ceramic capacitor of 30pF with one end grounded and the other connected to two projecting pins in a metal casing of the crystal oscillator. Oscillators and capacitors can be jointly packed in a case with three pins. This package is called a ceramic resonator. When a device is to be designed, it is a rule to place an oscillator nearer the MCU to avoid any interference on the lines on which the MCU is receiving a clock. This type of oscillator is used in devices that require great precision of time.

#### 2.2.1.2 RC-Oscillator

This type of oscillator is used in conditions where great precision is unnecessary. It depends on resistance, capacitance, supply voltage rate and working temperature to achieve a resonant frequency. The capacitor rating should be such that it controls noise and stability. When voltage is supplied, the oscillator begins to oscillate though it is unstable at first. It gradually begins to attain a stable period and amplitude. This simple and inevitable act can influence the MCUs performance and hence the MCU is placed in reset state during the process of the oscillator clock stabilizing. The oscillator used for this system is a crystal oscillator.

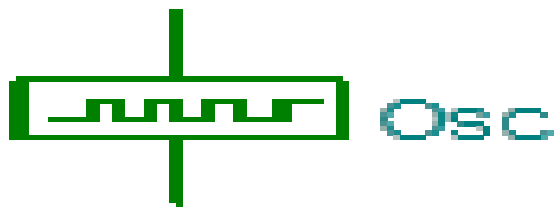

Fig. 2: Schematic representation of a clock oscillator

### 2.2.2. Programming

This is the art of communication with devices using an artificial language that can be used to define a sequence of instructions that can be processed and executed by such devices. There are two broad classifications of programming languages – the low-level and the high-level languages.

#### 2.2.2.1. Low-level Language

This type of language is written specifically for a particular type of micro-controller or microprocessor. This means that it cannot be used by another microcontroller. The instructions in this language are in mnemonics. This is called Assembly language.

#### 2.2.2.2. High Level Language

This type of language is formed from parts of natural language such as English. It is a high level of abstraction between what is asked by the computer and what the computer actually understands. It is easily understood by humans more easily than assembly languages. But, like the assembly language the computer cannot understand it. They therefore have to be translated into machine code the language the computer understands.

### 2.2.3. Machine Code

This is a sequence of carefully timed series of ON and OFF signals that can also be called high and low pulses or digital zeros and ones. The code usually represents numbers, data and instructions for manipulating those numbers and data.

## 3. System Design and Implementation

This section will discuss the design procedure and the real time implementation of the system. The working principle of the smart embedded PC uninterruptible power supply unit is visually explained in the schematic block diagram shown in Fig. 3. The inverter block, which is the central block in the design, does the inversion of a 12V DC to a 220V AC. This block provides the back-up power supply unit for the load in the case of power outage. The DC supply block is needed in order to charge the battery since the rechargeable batteries are not charged by AC voltages. Though not schematically shown, this block also powers various circuit components which would be extensively discussed in later sub-sections of this section. The switching circuit block does the automatic switching from AC mains to inverted DC power. The microcontroller block comprises a single microcontroller chip used for both interfacing with a conventional personal computer (software control) and for other circuit components control (hardware control). This system enables the automatic shutdown of the personal computer when the battery level falls below a designated safe voltage value. This occurs only when there is a power outage.

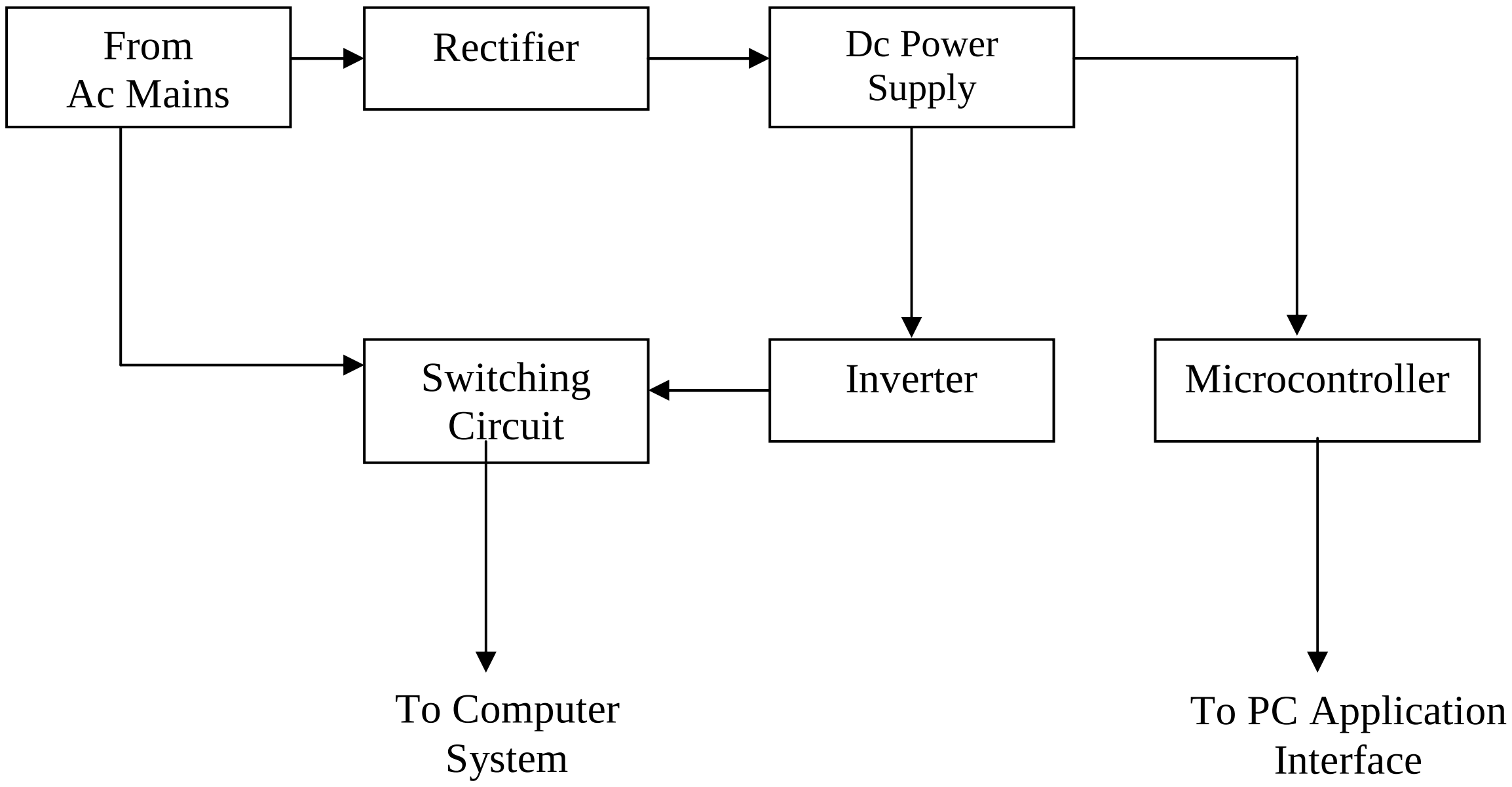

Fig. 3: Complete Block Diagram of 650VA Smart Embedded PC UPS

For simplicity of understanding, the design of this system has been sectioned into three (3) modules: (1) The power module, (2) The control module and (3) The inverter module.

## 3.1. The Power Module

This module is the circuit block that is necessary for the provision of regulated (stepped-down rectified) DC power supply (from AC mains) to circuit components (ICs, relays and microcontroller) for the charging of the rechargeable battery. The circuit diagram of this module is shown in Fig. 4. The input voltage from the mains is taken to be nearer to constant power supply from the power supply company (an approximate value of 220 – 240V AC). This circuit is protected by a fuse rated at 13A, 240V. The fuse as a protective device is supposed to break the circuit when there is a current flow above its rated value. The resistance of the fuse can be calculated from the equation.

$$R = \frac{V}{I} = \frac{240}{13} = 18.5\Omega \approx 20\Omega \tag{1}$$

### 3.1.1. Description of Components

The 240V AC is then passed to the primary winding of a 240/15V step-down transformer (i.e. secondary voltage $\leq$ primary voltage). The choice of this transformer would be explained in due course. However the transformer internal analysis is thus, the ration of Primary winding over the secondary winding as:

$$\frac{V_2}{V_1} = \frac{N1}{N2} = \frac{240}{15} = 16 \tag{2}$$

$$\frac{N2}{N1} = 10^{-1}; \frac{N1}{N2} = 10 \tag{3}$$

$$\frac{V_2}{V_1} = \frac{I_1}{1_2} \qquad (4)$$

In the peak secondary voltage $V_P$, the ratio $N_1/V_1$ is called the forms-per-volt-ratio. The primary and secondary of the transformer have the same turns-per-volt-ratio.

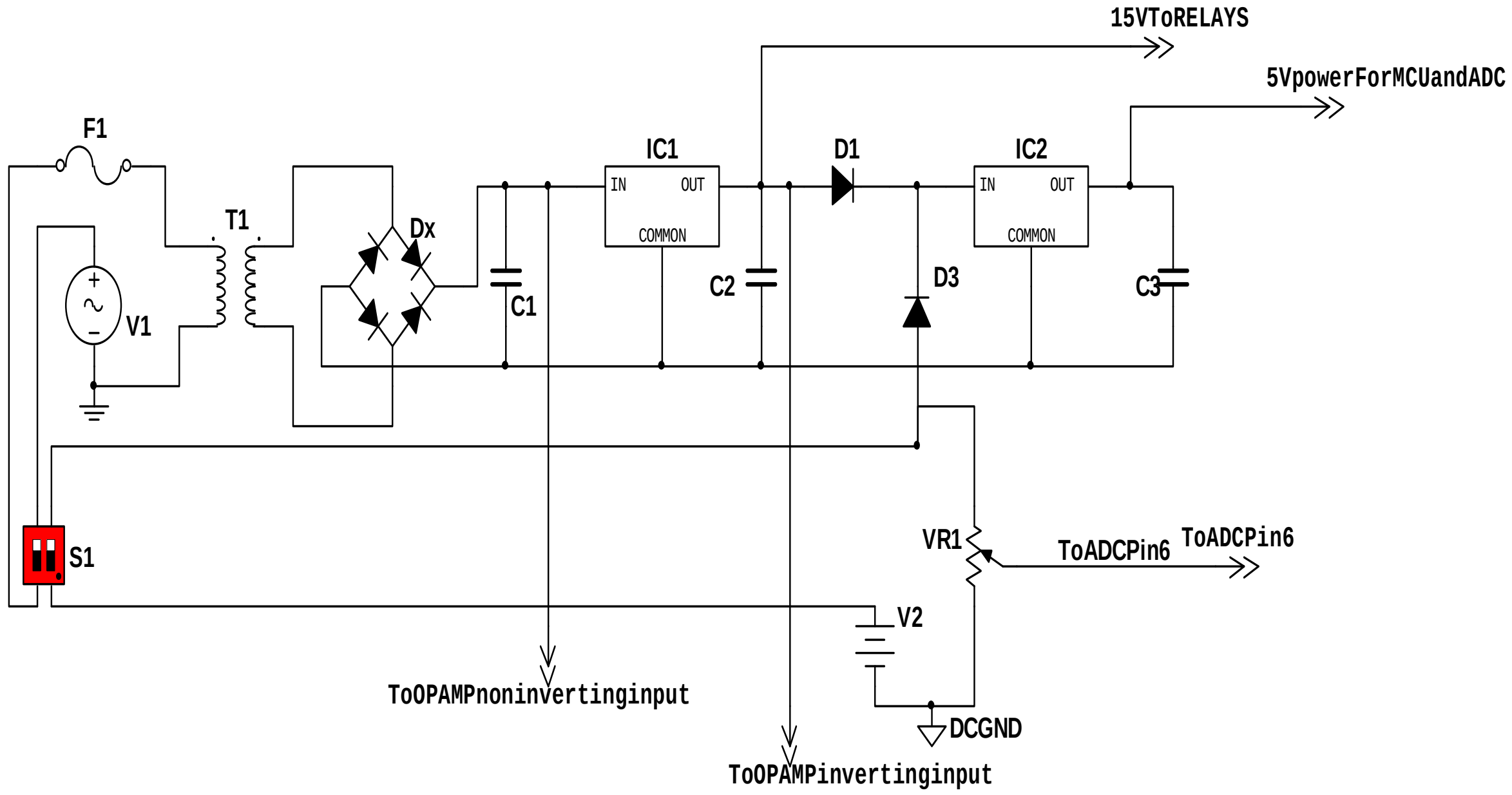

Fig. 4: Complete power module circuit diagram

### 3.1.1.2. Voltage Regulators (LM7815 and LM78L05)

These two voltage regulators are used to give a constant DC voltage of 15V (LM7815) and 5V LM78L05. They act as stabilizers due to the fact that the circuit components are to run on DC voltage that contains negligible or no pulsations at all. These regulators give an unvarying output. The LM7815 uses a heat sink due to its nature to heat up. The LM78L05 however does not need a heat sink. Both the two regulators have a maximum current drawn of 1A each. The LM7815 gives an output of 15V that is fed into the comparator (LM741), though due to configurations it is not directly used as a reference voltage. The two relays RLAI and RLA2 also feed from this terminal. The LM78L05 gives an output of 5V that is fed to the microcontroller unit. This terminal must at all times have an output of 5V either from the rectified power or the battery terminal because the microcontroller oversees the general control of the whole circuit and must always be powered. This regulator is fed by a joint from two diodes (IN4001) which prevent a flow back of current and are the alternating sources of voltage to the regulator.

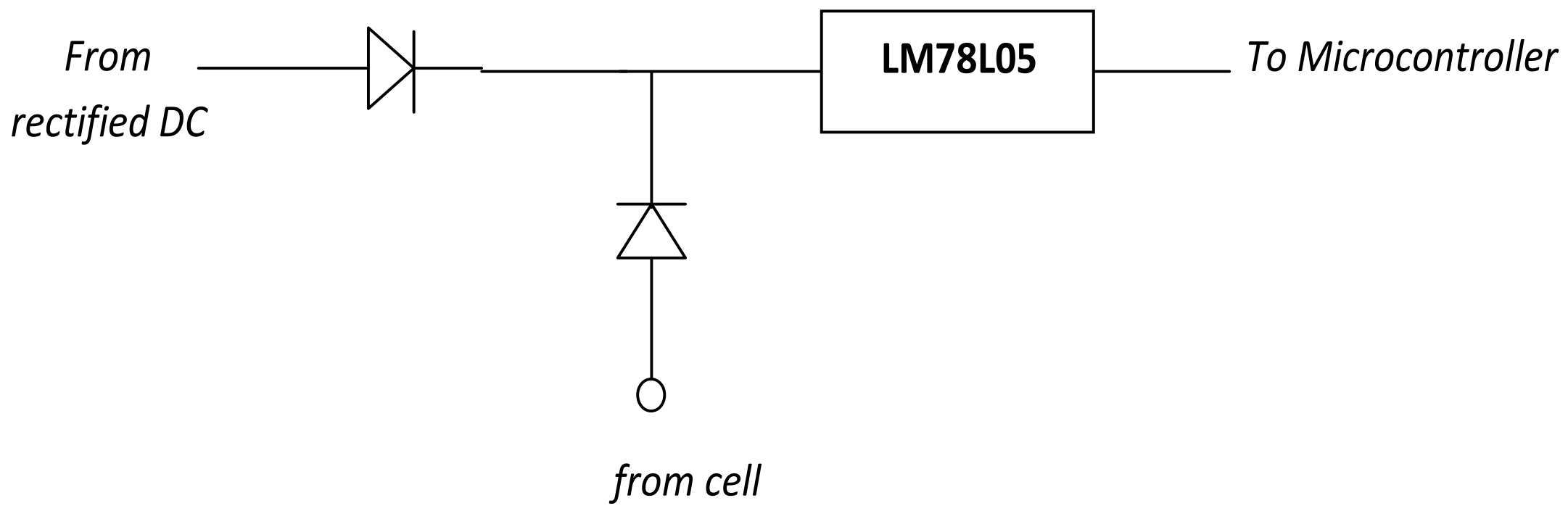

Fig. 5: Connection of the 5V voltage regulator (LM78L05)

### 3.1.1.3. Other Components

Other components are the capacitors, resistors and diodes used in the power supply stage. The various capacitors in this module act as filters with various ratings and their respective bleeding resistors in this case a 300Ω.

TABLE 1: Ratings of capacitors

| Component | Farad | Voltage | State |
|---|---|---|---|
| C1 | 2220µF | 10V | Polarized |
| C2 | 470µF | 16V | Polarized |
| C3 | 10µF | 25V | Polarized |

$C_1$ is higher because we have a greater level of pulsation at the point of immediate rectification.

Diodes: The two diodes used in this module are basic diodes and all they do is to prevent a flow back and a constant inflow to the Voltage regulator. They have the following configurations.

Table 2: Ratings of Diodes

| Component | Type |
|---|---|
| D1 | IN4001 |
| D2 | IN4001 |

### 3.1.1.4. The Rectifier Component

Rectification of AC to DC is achieved by a full-wave bridge (FWB) rectifier configuration. The FWB has twice the efficiency of the half-ware and its output is equally twice. A FWB as opposed to a two-diode full-wave rectifier when used can result in the same DC voltage but the transformer used with the two-diode full-wave rectifier must have higher turns ratio (N1/N2). This implies that with the bridge rectifier, fewer turns of wire are needed in the transformer. This results in a smaller, lighter and cheaper transformer. This benefit far more exceeds the cost of the additional two (2) diodes that constitute the bridge rectifier.

The full-wave bridge rectifier: This is a setup of four (4) diodes (1N4001) configured in a manner that the pulsating AC voltage they encounter is transformed to a positive amplitude only that is the negative amplitudes are eliminated.

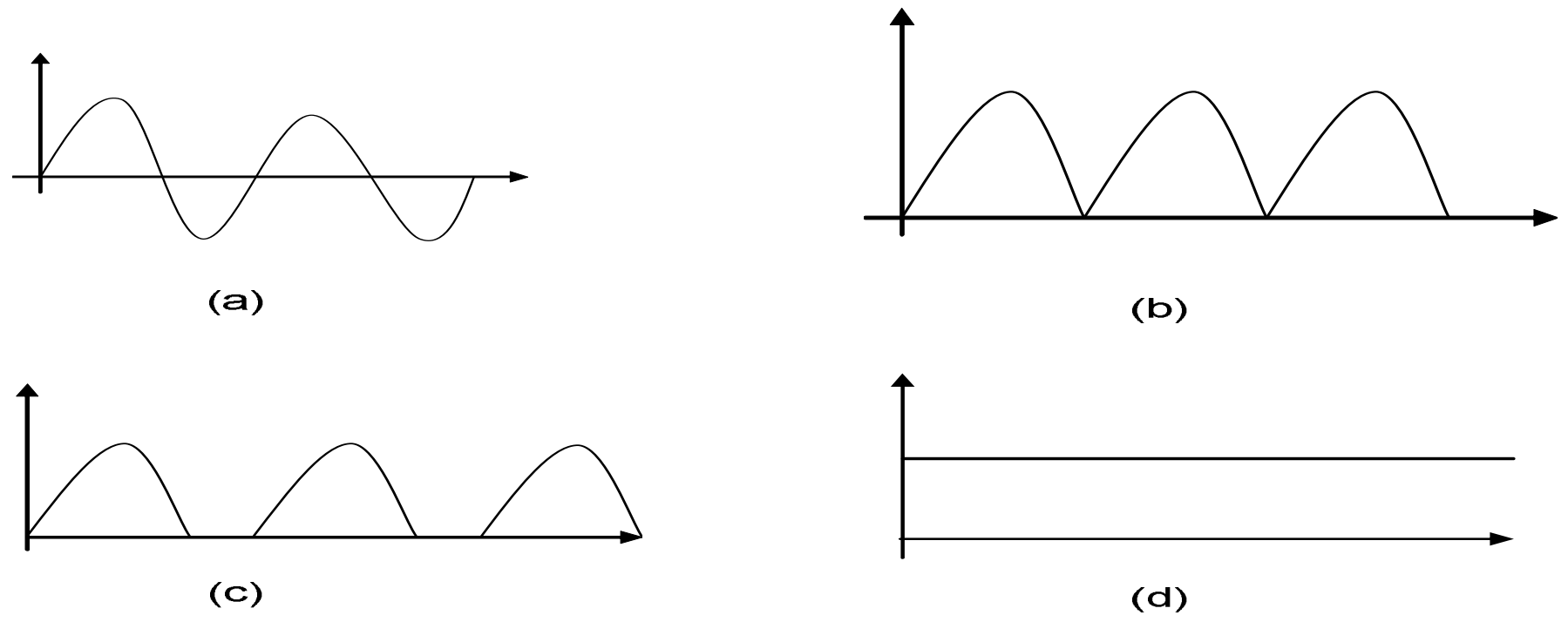

Fig. 6: (a) Waveform of AC output, (b) Waveform of improperly filtered DC output, (c) Waveform of Unfiltered DC output and (d) Completely filtered DC output.

From the diagram above, Fig (a) shows a pulsating AC, Fig. (b) Shows a pulsating DC (improperly filtered), Fig. (c) Shows a pulsating DC (without a filter) and Fig. (d) Shows a DC that has undergone filtering and hence is no longer pulsating. That is all ripples have been smoothened.

## 3.2. The Control Module

The control module is the driver of the whole circuit. It determines different rates like the charging value of the battery, the switch voltage for the relays and the battery safe voltage value. The components used in this module are outlined in the subsequent sub-sections.

### 3.2.1. The Microcontroller

The main controller is handled by the microcontroller (AT89C2051) that runs on a crystal of 12MHz frequency. The microcontroller has twenty (20) pins. Pin 1 is the reset pin and the reset circuit is shown below in Fig. 7.

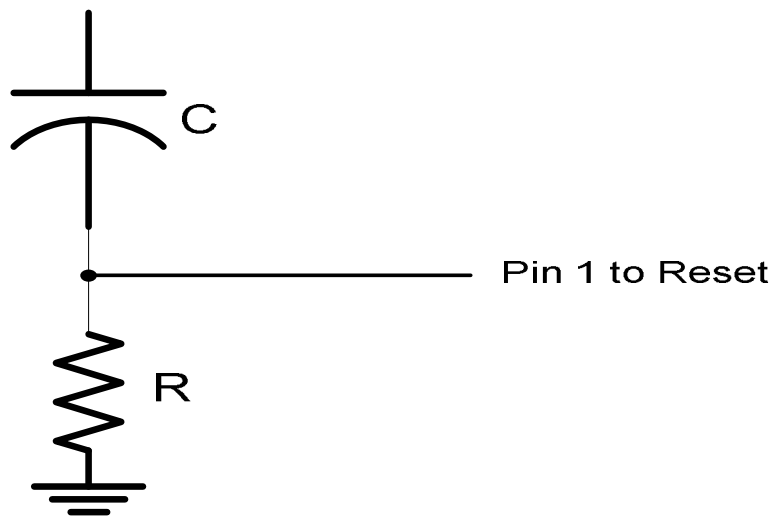

Fig. 7: Configuration of Reset Circuit for MCU Pin 1

The reset pin sets the MCU to a neutral point so as not to be influenced by the initial unstable nature of the oscillator attached to the MCU at pins 4 and 5. Pin 2 through 9 are used for the parallel port interface. This is the soft control that is handled by the computer operating system. The software (***Xmart 1.0***) that performs the soft control is written in Microsoft Visual C++ high level language. This language is chosen due to its proper management of hardware devices. The MCU however runs on Assembly language (Low level language). Due to page limits, we are not able to put all the codes used for the system. We were only able to provide the codes written for the MCU in an assembly language, and leaving the codes for the software application (*Xmart 1.0)* written in Microsoft visual C++. Rather, we have described the algorithm with a flow chart (both for ***Xmart 1.0*** and the MCU's Assembly language). Though, interested readers can contact us directly for the source codes.

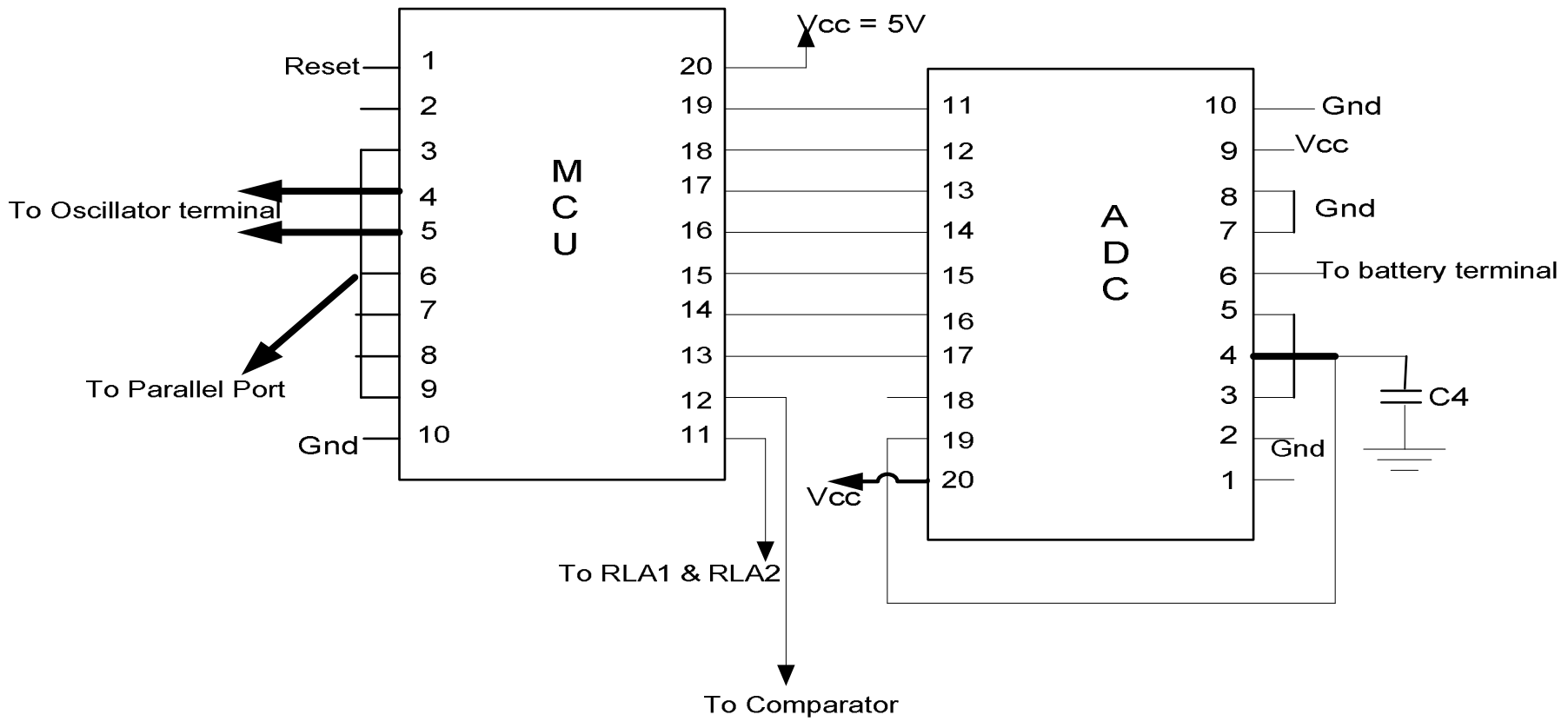

Fig. 8: Diagram of MCU interfaced with ADC

In the diagram above (Fig. 8) Showing the interfacing between MCU with ADC, Pins 13 – 19 are connected to the outputs of an ADC (Analog-to-Digital Converter), Pin 12 is linked to the comparator, Pin 11 is connected to the relays RLA1 and RLA2, Pin 20 is the VCC and Pin 10 is connected to the ground.

### 3.2.2. The Comparator (LM741)

This comparator has the duty of checking for an AC output that can conveniently power the entire circuit. The reference voltage is set by the variable resistors and a benchmark of 180V AC is fixed. This benchmark was chosen due to a test carried to observe that the motherboard of most computers would call for a restart at this AC value. The comparator therefore switches the relays to the battery mode when AC drops to 180V or below.

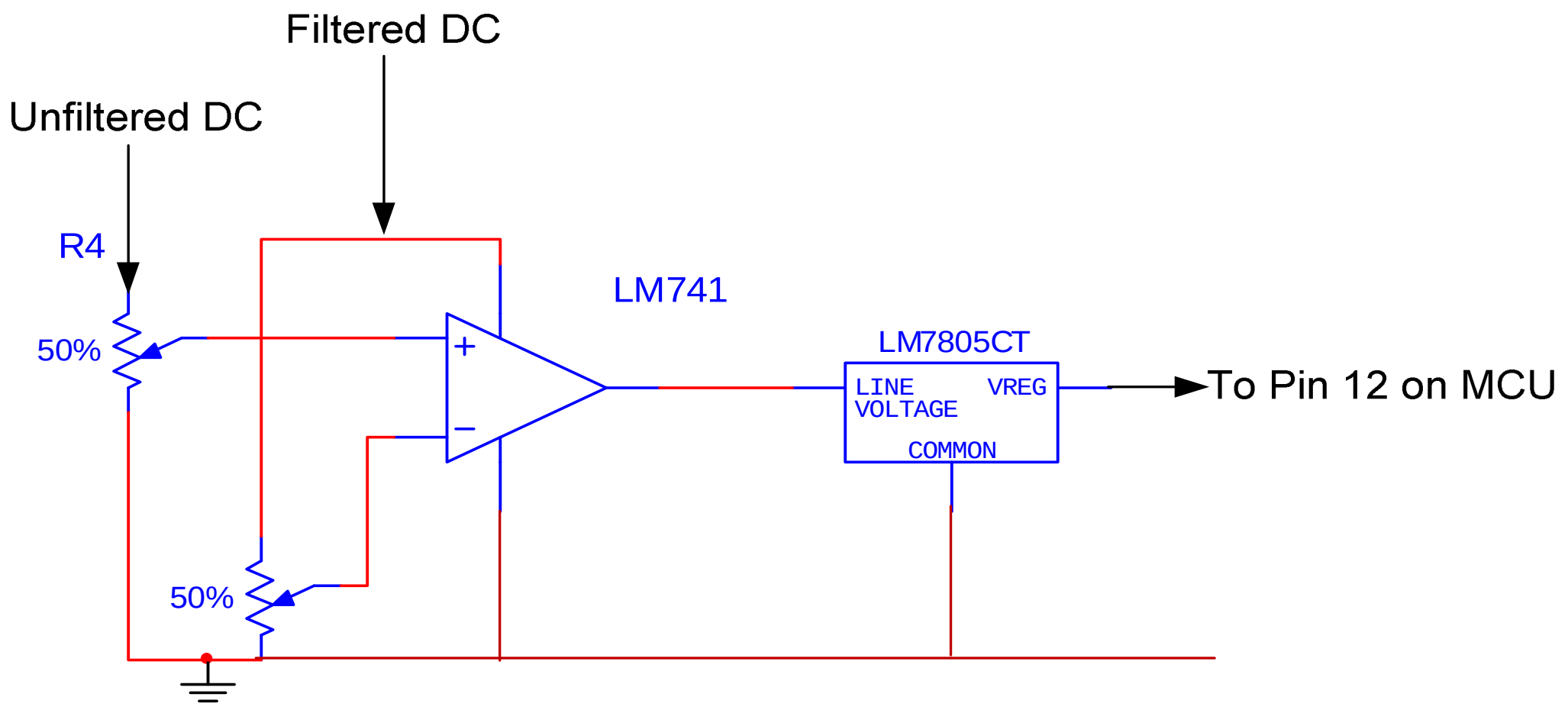

Fig. 9: Configuration of Voltage Comparator

The output of the comparator is sent to a voltage regulator LM78L05 (see Fig. 9) whose output is sent to pin 12 of the microcontroller. This activation of the soft control automatically. Note however that the software control can also be manually activated to shut down the system.

### 3.2.3. The Relays

The relays are simple automatic switches. Their configuration is given below:

| Component | Volt | Volt type | No. of poles |
|---|---|---|---|
| RLA 1 | 12V | DC | 1 |
| RLA 2 | 12V | DC | 2 |

RLA1 switches by impulses from the microcontroller. It switches between the charging voltage (15V) point and a dumb terminal. The MCU tells RLA1 when to begin to charge the battery. This relay is never powered when there is a power outage. RLA2 has 2 terminals, which switches between the inverter and the mains supply. It also switches the battery to a dumb terminal when the inverter does not require it.

## 3.3. The Inverter Module

This module is the segment of circuitry that is responsible for the conversion of 12 volts DC from battery supply to 220 volts AC that supply backup power to the load. This module consists primarily of an oscillator circuit shown in Fig. 10 that causes two sets of MOSFETs to switch alternately at a frequency of 50Hz (in compliance with present power supply frequency), and a step-up transformer that steps up the voltage to 220 volts AC.

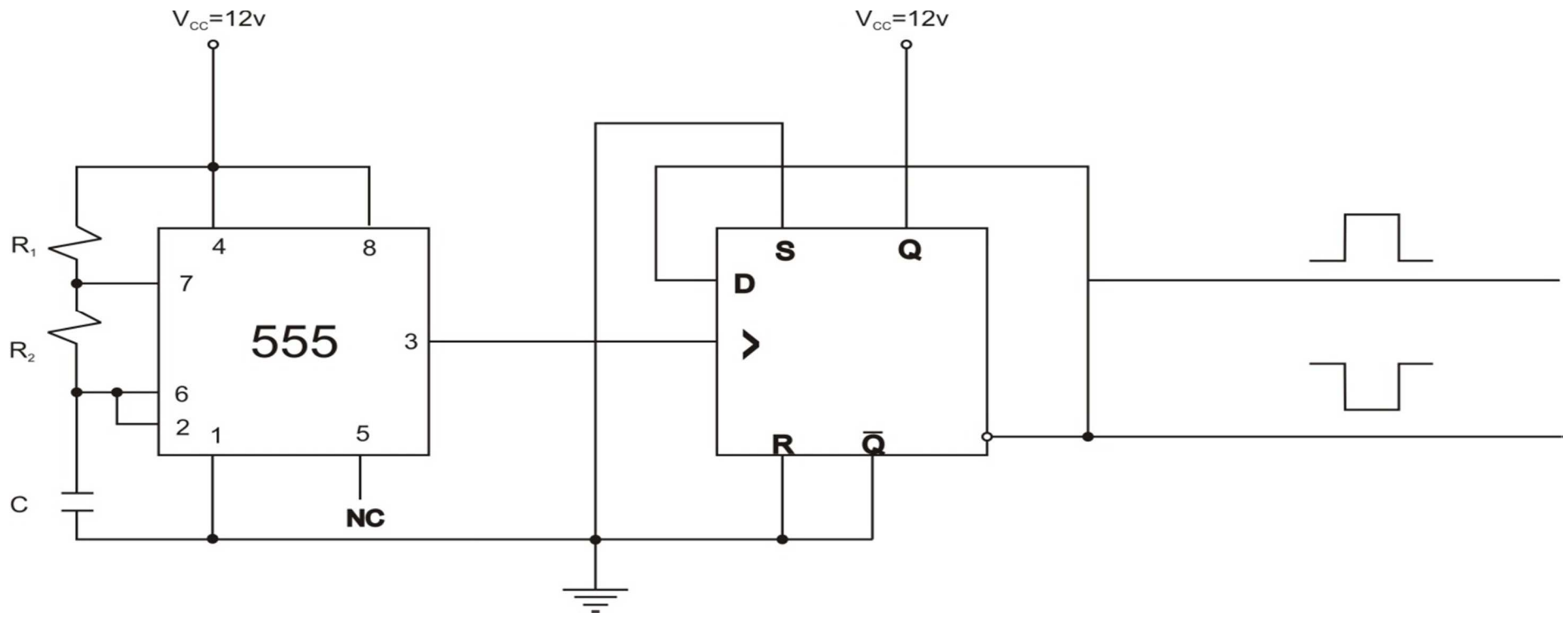

Fig. 10: Bistable Oscillator Circuit

The oscillator circuit was implemented with the aid of a 555 timer configured to operate as an astable multivibrator with its output fed into a D flip-flop producing two outputs (Q and $\overline{Q}$) of inverted waveforms. The frequency of oscillation of the 555 timer was set with the formula:

$$F = 1.44 / (R_1 + 2 R_2) C \qquad (5)$$

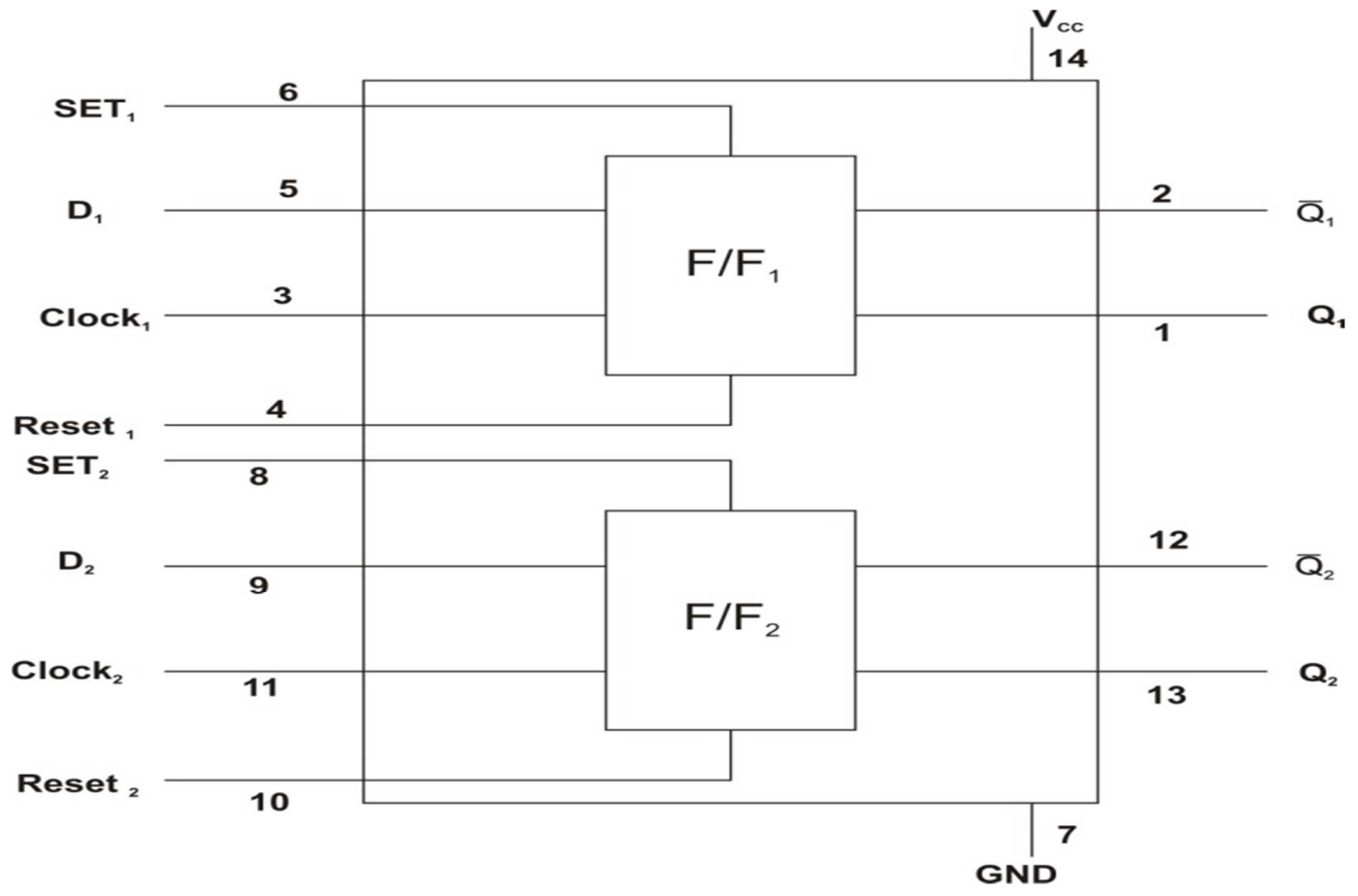

Fig. 11: Pin configuration of the CD4013 IC

$R_1$ was chosen to be 4.7kΩ and C was chosen to be 1µF. $R_2$ was chosen to be a 100kΩ variable resistor and it was varied while the frequency of the timer was being measured (with a frequency meter) until a 100Hz frequency was achieved. The output of the timer was then passed through the D Flip-flop that divides the frequency in two (50Hz) while providing two outputs that are the inverse of each other. The two outputs of the flip-flop were fed into the gates of two sets of MOSFETs that act as alternate switches while the transformer completed the circuit by stepping up the voltage to the desired level. Care was taken, for proper saturation of the FETs (MOSFETs), to ensure that the gate voltage was approximately equal to the drain voltage. This was achieved by powering the oscillator circuit from the same source of power as that of the FETs (the battery).

In this module, a conventional 555 timer which is an 8-pin IC package was used as the first stage of the oscillation circuit. The D flip-flop used was a CD4013 IC that consists of dual flip-flops in one package. Only one of the flip-flops, the upper part of the package was used, with the set, reset, clock and data pins of the other flip-flop in the package grounded. The pin configuration of this is shown in Fig. 11. The transformer used is a 12-0-12/240V step-up transformer wound to handle a 600W load. The FETs used (Z44) had current rating higher than that of the battery (12V 17Ah) to ensure that they don't burn up on powering them, since they draw up maximum current from the battery. Two IN5408 diodes were also placed across the FETs for protection against backward current flow as shown in Fig. 12.

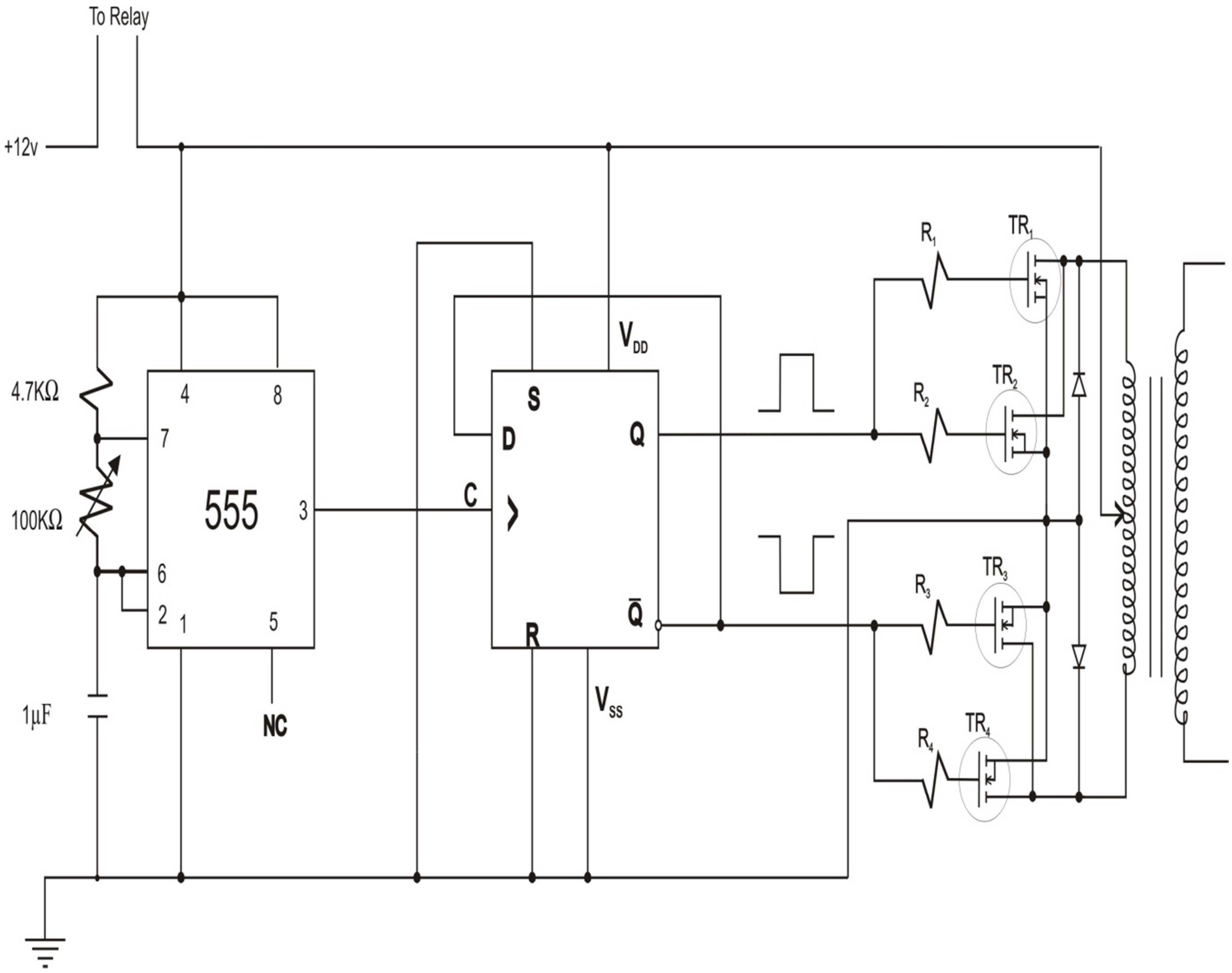

Fig. 12: The Inverter Module

### 3.4. Programming and Procedural Flow

The control module offers a series of flow to determine various conditions of operation. Below is a summarized flowchart of the module

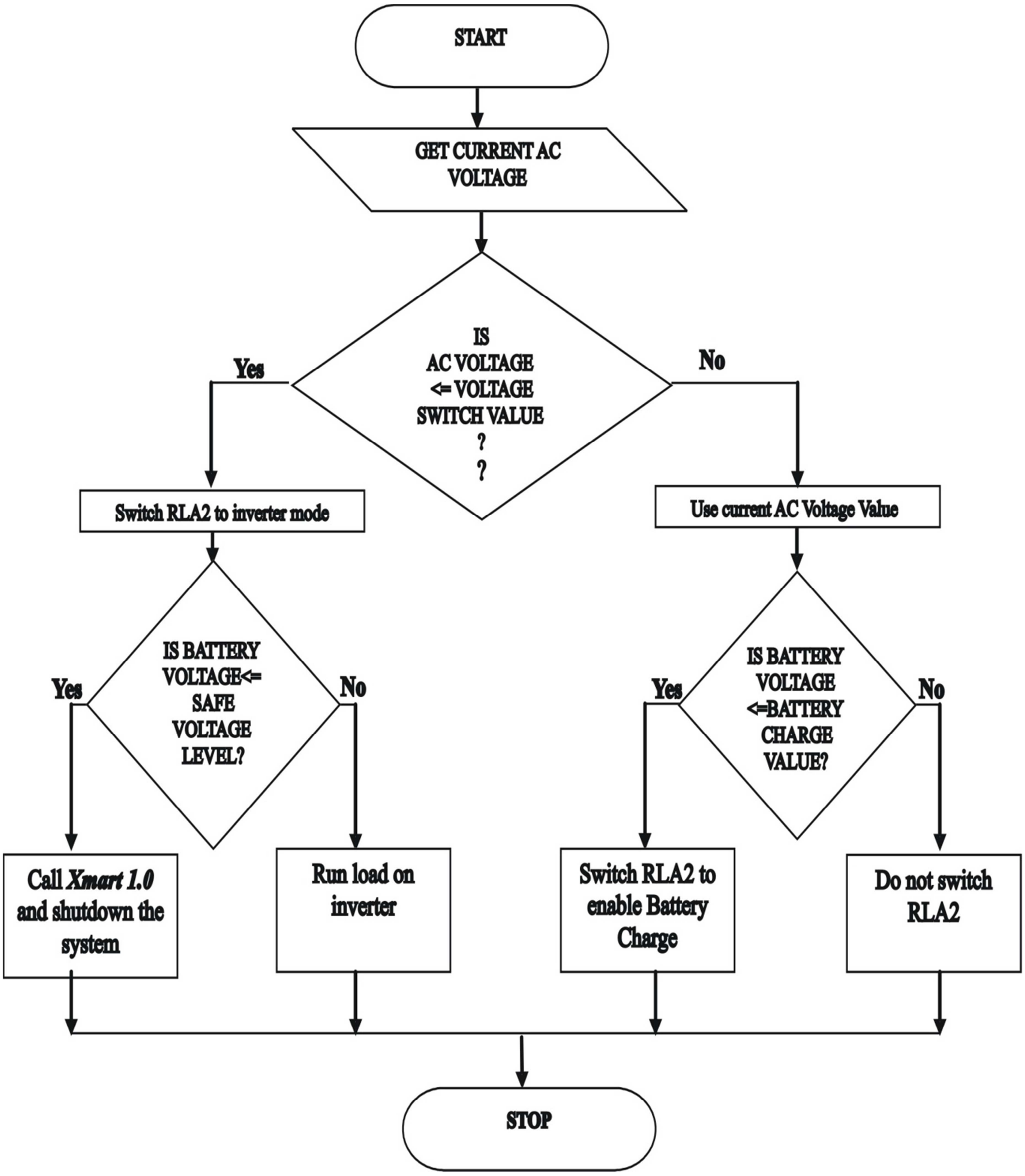

Fig. 13: Summarized flowchart of the procedural flow of the control module

Also, the visual displays of XMART 1.0 under development and after development are shown in Fig. 14 and Fig. 15 respectively.

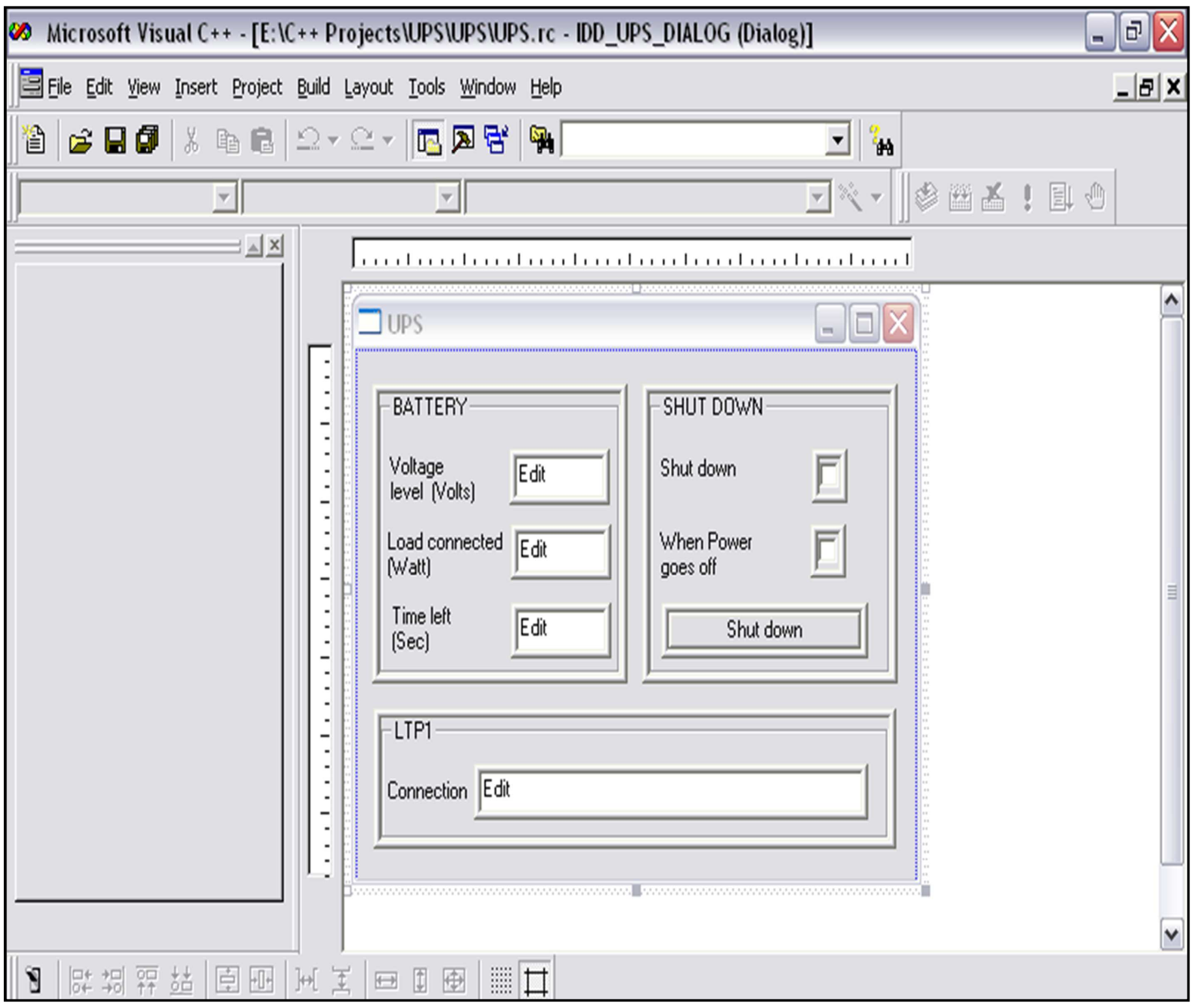

Fig. 14: Developer Environment for Microsoft Visual C++ showing Xmart 1.0 under development

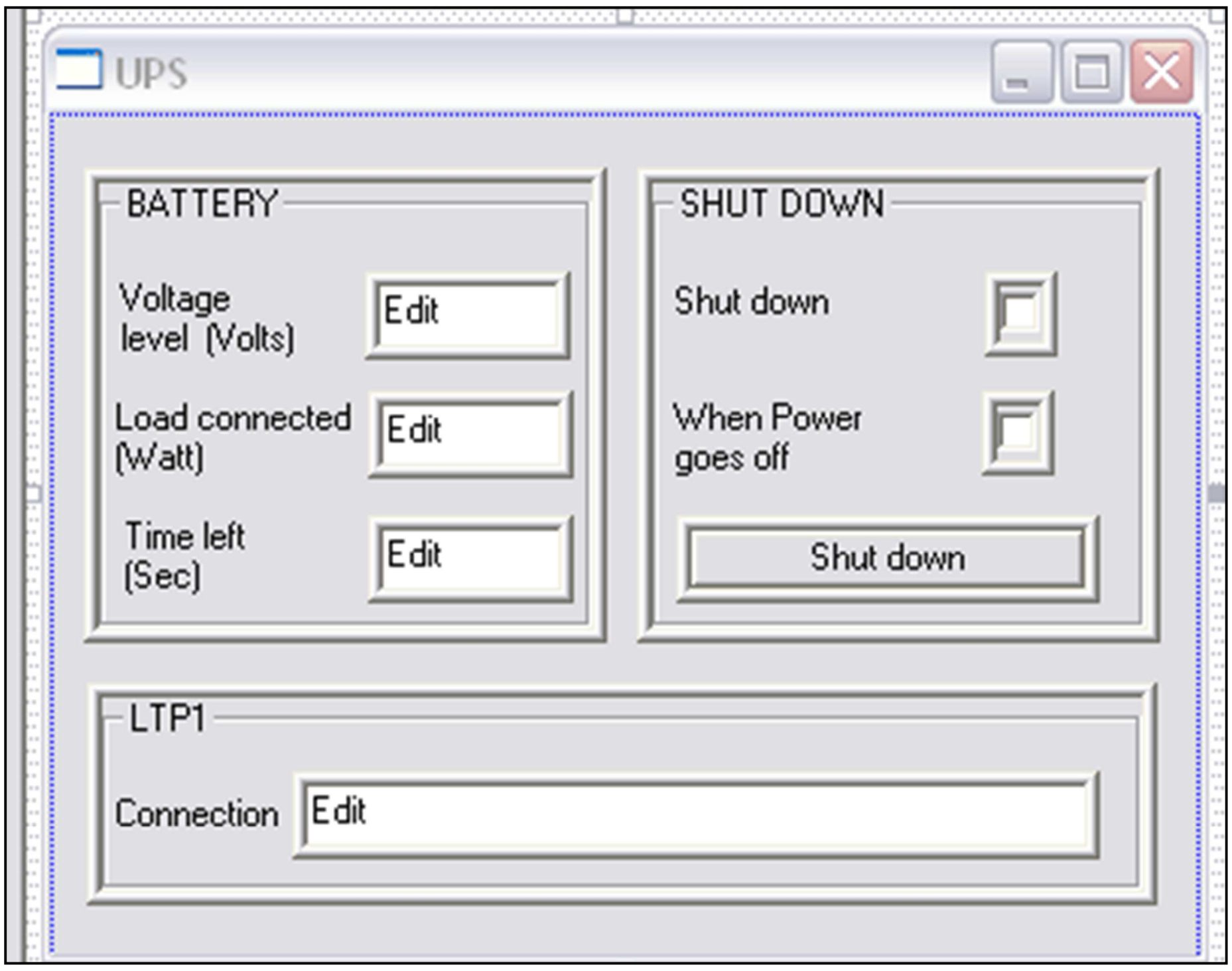

Fig. 15: Visual display of Xmart 1.0 after development

### 3.5. Design Specification

Since this project is designed to handle a conventional computer system unit, specifications had to be sought to ensure that no undue outputs or designs were gotten or made respectively. Below (Table 3) is the basic power consumption rating of a personal computer system.

**Table 3: Computer Power Rating Specification**

| | COMPUTER SYSTEM PARTS | WATTS (W) |
|---|---|---|
| 1 | MOTHERBOARD (without CPU or RAM) | 30 |
| 2 | 550MHz PENTIUM III | 30 |
| 3 | 7200RPM IDE HARD DISK DRIVE (HDD) | 15 |
| 4 | RAM (128MB) | 10 |
| 5 | 50x CD-ROM DRIVE | 25 |
| 6 | NETWORK INTERFACE CARD (NIC) | 4 |
| 7 | FLOPPY DISK DRIVE (FDD) | 5 |
| 8 | PERIPHERAL COMPONENT INTERCONNECT (PCI) CARD | 5 |
| 9 | ACCELERATED GRAPHIC PORT (AGP) CARD | 30 |
| 10 | VISUAL DISPLAY UNIT (MONITOR) | 330 |
| | **APPROXIMATE POWER CONSUMPTION** | **484** |

The design that this UPS incorporates therefore has the following specifications.

**Table 4: Specifications for System Design**

| | |
|---|---|
| OUTPUT POWER | 650VA |
| OUTPUT FREQUENCY | 50HZ |
| OUTPUT VOLTAGE | 20V |
| INPUT VOLTAGE | 12V DC |

This system creatively combines the circuitry of a conventional DC Power Supply Unit (Power Pack) to achieve an exact voltage distribution to all components of the computer system. The circuitry of the power pack if altered would enhance the compactness but that is beyond the scope of this design.

# 4. TESTING, RESULTS AND PROBLEM ENCOUNTERED

This project necessitated some very vital tests to ascertain the various required inputs and generate the proposed output. The following experimental tests were carried out and are documented below.

## 4.1. Voltage Switch Test

It was noticed that at a particular AC voltage the system unit was automatically restarted. By the use of a dimmer and test Multimeter, the voltage was varied from an initial value of 220V. It was noticed that at a range of 180-150V AC the system unit restarted. This point was hence made to be the voltage value point at which the UPS is supposed to switch the power supply to the inverter. This test produced a range of 180V AC and below. It is assumed that at this range there is a cut in the power supply of the main.

## 4.2. Battery Safe Value Test

This test was carried out to determine the value of battery voltage at which it would be unsafe to continue running the system on the inverter as a source. The purpose of this test is also seen in the methodology employed by the MCU, whereby it sends a signal to the parallel port to activate *Xmart 1.0* (the software application that is configured for the UPS) and shuts down the system within a given time (*Xmart 1.0* shutdown time frame) interval to avoid completely running down the battery. This value was set to 6.0V.

## 4.3. Battery Charge Value

This value was set to determine the voltage at which the MCU could switch the relay that enables the battery to begin charging. It was chosen to be 11.5V and the full charge value to be 13.5V this is the specified charge value.

## 4.4. Time Interval Test

This test was to determine the rate at which the battery runs down on rated load. The results from this test were to help assume a time frame for which the UPS can sustain the system on battery power. It however was not feasible.

## 4.5. XMART 1.0 Shutdown Time Frame

When the MCU senses that the battery has gotten to the safe value, it calls up the software application (*Xmart 1.0*) and it hands over control to the application. *Xmart 1.0* gives the user a time frame of sixty (60) seconds to quickly save documents after which it calls the shutdown function of the operating system and shuts down the system.

## 4.4. Problem Encountered

1. One of the main problems encountered was in the control module. The initial MCU had an internal comparator but it could not handle a value of 12V. This led to the inclusion of an ADC to be able to convert the 12V from the battery to a rectifiable pulse for the MCU.
2. Compatibility was also a major issue, because the purpose of the project would have been defeated if its bulky nature was not better than what exists. This desire to make the project compact actually led to the use of an MCU and DC relays. Also the battery terminals were externally placed to reduce the weight of the complete model.

3. The software control implementation was supposed to use a USB interface due to its convenient internal placement. However, a parallel port external interface was chosen due to the modernized and highly technical nature of a USB port in hardware interfacing.
4. The application software written in Visual C++ would have been done in Visual Basic since it is relatively simple. It however does not handle hardware implementation and so could not be used. Other high level languages are too advanced for the project at hand.

## 5. Conclusions

It can be concluded that the sole aim of carrying out the design, analysis and implementation of *a smart embedded personal computer uninterrupted power supply system* was achieved, in that the aim was to develop a cheap, affordable, reliable and efficient smart embedded system, which was successfully realized at the end of the design process. The whole concept of the system cuts across the hardware implementation and software implementation. The power module generated an output that conveniently powers a personal computer and the control module do the master channeling of device outputs and inputs though they are controlled mainly by the assembly code on which the microcontroller runs on. However similar implementations existed before now and were called internal UPS. Unique to this design however is the principle behind the control of the module, whereby a 5V microcontroller has to read a source of 12V (DC).

## Acknowledgements

The authors would like to thank Col. Muhammed Sani Bello (RTD), OON, Vice Chairman of MTN Nigeria Communications Limited for supporting the research.

**Author**

**Engr. Adamu Murtala Zungeru** received his B.Eng. degree in Electrical and Computer Engineering from the Federal University of Technology (FUT) Minna, Nigeria in 2004, and M.Sc. degree in Electronic and Telecommunication Engineering from the Ahmadu Bello University (ABU) Zaria, Nigeria in 2009. He is a Lecturer Two (LII) at the Federal University of Technology Minna, Nigeria in 2005-till date. He is a registered Engineer with the Council for the Regulation of Engineering in Nigeria (COREN), Member of the Institute of Electrical and Electronics Engineers (IEEE), and a professional Member of the Association for Computing Machinery (ACM). He is currently a PhD candidate in the department of Electrical and Electronic Engineering at the University of Nottingham. His research interests are in the fields of automation, Security, swarm intelligence, routing, wireless sensor networks, energy harvesting, and energy management.

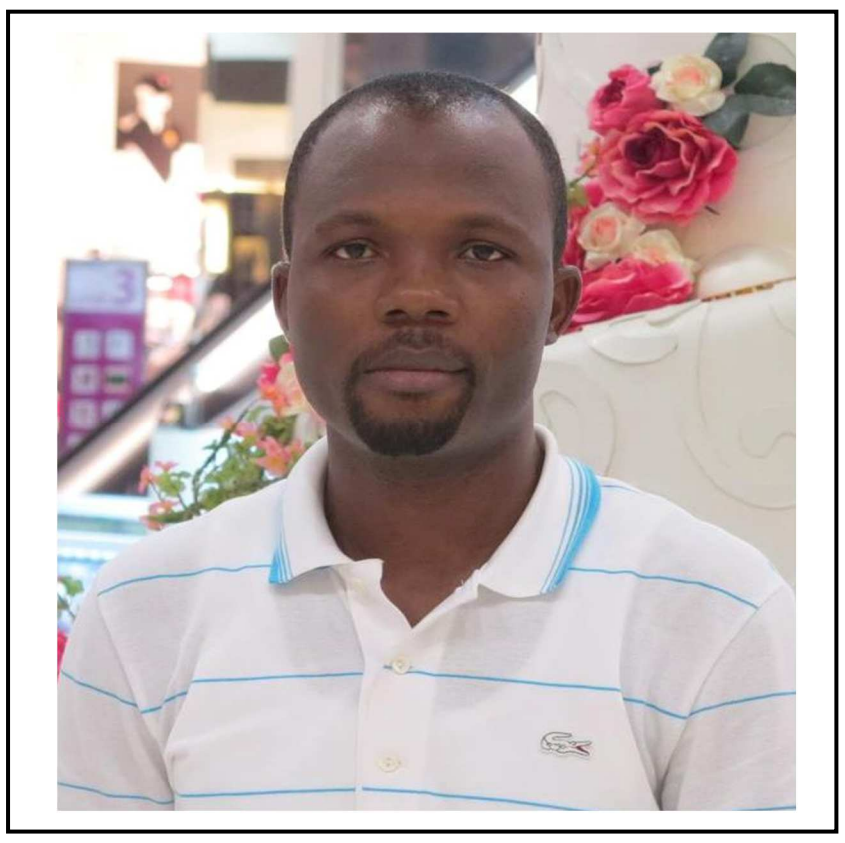

Appendix A: Codes Written for the MCU in Assembly Language

```
;=============================================================
;=============================================================
;            UNINTERRUPTED POWER SUPPLY
;==============================================================
;==============================================================
;          DEFINATIONS
;==============================================================
;==============================================================
;REGISTERS
Ctr         Equ 02h
;BIT MEMORY
ADCReg    Equ 20h
PPReg       Equ 21h
;PORT
ADCPort     Equ P1
PPort       Equ P3
;BIT
ADCReg0   Equ 00h
ADCReg1   Equ 01h
ADCReg2   Equ 02h
ADCReg3   Equ 03h
ADCReg4   Equ 04h
ADCReg5   Equ 05h
ADCReg6   Equ 06h
ADCReg7   Equ 07h
PPReg0    Equ 08h
Charger     Equ P3.7
Mains       Equ P1.0
;=============================================================
;=============================================================
;    VECTOR ADDRESSES
;=============================================================
;=============================================================
;=============================================================
    Org 0000h        ;RESET VECTOR ADDRESS
    ljmp Start       ;Jump to start of program
;=============================================================
    Org 0003h        ;EXTERNAL INTERRUPT0 VECTOR ADDRESS
    reti             ;Return from Interrupt
;=============================================================
    Org 0Bh          ;TIMER0 INTERRUPT VECTOR ADDRESS
            acall Timer
    reti
;=============================================================
    Org 13h          ;EXTERNAL INTERRUPT1 VECTOR ADDRESS
    reti
;=============================================================
    Org 1Bh          ;TIMER1 INTERRUPT VECTOR ADDRESS
    reti
;=============================================================
    Org 23h          ;SERIAL INTERRUPT VECTOR ADDRESS
    reti             ;Not used
;=============================================================
    Org 30h          ;Program starts here
Start:
    mov SP,#40h     ;Stack Pionter intialized
```

```
    clr RS0         ;Bank0 selected
    clr RS1
            mov PPort,#255
            mov ADCPort,#255;Initialising ADCPort as input port
            setb Mains
            mov Ctr,#8
            mov TMOD,#17    ;Timer0 (16bit Timers)
            mov TH0,#11    ;Timer0 reload value= 55535
    mov TL0,#219
    setb ET0        ;Timer0 Interrupt enabled
            setb TR0        ;Start Timer0
    setb EA                          ;Global interrupt enabled
            jmp $           ;Wait
;==========================================================
Timer:
    clr TR0         ;Stop Timer0
            mov TH0,#11    ;Timer0 reload value= 3035
    mov TL0,#219
    setb TR0        ;Start Timer0
                        djnz Ctr, EndTimer
            mov Ctr,#8
            mov ADCReg,ADCPort
            setb ADCReg0
            acall ADConversion
    acall ChargeBattery
            acall UpdatePPort
EndTimer:
    ret
;==========================================================
ADConversion:
            mov dptr,#PPData
            mov A,ADCReg
            movc A,@A+dptr
            mov PPReg,A
   ret
;==========================================================
ChargeBattery:
    mov A,ADCReg
            cjne A, #10,ChargeBat
            cjne A, #12,StopCharging
    ret
ChargeBat:
    clr Charger
            ret
StopCharging:
    setb Charger
            ret
;==========================================================
UpdatePPort:
    jb Mains,lMains
    clr PPReg0
            mov C,Charger
    mov PPort,PPReg
            mov Charger,C
    ret
lMains:
            mov C,Charger
    mov PPort,#255
            mov Charger,C
```

```
        ret
;==========================================================
;==========================================================
PPData:  db 243,243,243,243,243
    db 242,242,242,242,242
                db 241,241,241,241,241
                db 240,240,240,240,240
                db 239,239,239,239,239
                db 238,238,238,238,238
                db 237,237,237,237,237
                db 236,236,236,236,236
                db 235,235,235,235,235
                db 234,234,234,234,234
                db 233,233,233,233,233
                db 232,232,232,232,232
                db 231,231,231,231,231
                db 230,230,230,230,230
                db 229,229,229,229,229
                db 228,228,228,228,228
                db 227,227,227,227,227
                db 226,226,226,226,226
                db 225,225,225,225,225
                db 224,224,224,224,224
                db 223,223,223,223,223
                db 222,222,222,222,222
                db 221,221,221,221,221
                db 220,220,220,220,220
                db 219,219,219,219,219
                db 218,218,218,218,218
                db 217,217,217,217,217
                db 216,216,216,216,216
                db 215,215,215,215,215
                db 214,214,214,214,214
                db 213,213,213,213,213
                db 212,212,212,212,212
                db 211,211,211,211,211
                db 210,210,210,210,210
                db 209,209,209,209,209
                db 208,208,208,208,208
                db 207,207,207,207,207
                db 206,206,206,206,206
                db 205,205,205,205,205
                db 204,204,204,204,204
                db 203,203,203,203,203
                db 202,202,202,202,202
                db 201,201,201,201,201
                db 200,200,200,200,200
                db 199,199,199,199,199
                db 198,198,198,198,199
                db 197,197,197,197,197
                db 196,196,196,196,196
                db 195,195,195,195,195
                db 194,194,194,194,194
                db 193,193,193,193,193
;==========================================================
;==========================================================
end
```